\documentclass[epj]{svjour}
\usepackage{graphicx}
\usepackage{amsmath}

\begin{document}
\title{Large-deviation properties of resilience of transportation networks}
\author{A. K. Hartmann\inst{1}}
\institute{
Institut f\"ur Physik,
Carl von Ossietzky Universit\"at Oldenburg,
26111 Oldenburg, Germany\\
 \email{a.hartmann@uni-oldenburg.de} }

\date{Received: date / Revised version: date}

\abstract{
Distributions of the resilience of transport networks are studied
numerically, in particular the large-deviation tails. 
Thus, not only typical quantities like average or variance but the 
distributions
over the (almost) full support can be studied.
For a proof of principle, 
a simple transport model based on the edge-betweenness and 
three abstract yet widely studied random network ensembles are considered here:
Erd\H{o}s-R\'enyi random networks 
with finite connectivity, small world networks and spatial networks
embedded in a two-dimensional plane.
Using specific numerical large-deviation techniques, probability densities 
as small as $10^{-80}$ are obtained here.
This allows to study typical but also 
the most and the least resilient networks. The resulting 
distributions fulfill the mathematical large-deviation
principle, i.e., can be well described by rate functions in the
thermodynamic limit. The analysis of the limiting rate function reveals that
the resilience follows an exponential distribution almost everywhere. An analysis
of the structure of the network shows that
the most-resilient networks can be obtained, as a rule
of thumb, by minimizing the diameter of a network. Also, trivially,
by including more links a network can typically be made more resilient.
On the other hand, 
the least-resilient
networks are very rare and characterized by one (or few) small core(s) to which
all other nodes are connected. In total, the spatial network ensemble turns
out to be most suitable for obtaining and studying resilience of real
mostly finite-dimensional networks. Studying this ensemble in combination
with the presented large-deviation approach for more realistic, in particular dynamic
transport networks appears to be very promising.}

\maketitle

%
%

\section{Introduction}
\label{sec:intro}

Transportation networks 
\cite{albert2002review,newman2003review,newman_book2006,dorogovtsev_book2006,cohen_book2010,barrat_book2012}, 
like computer networks, railway systems, water pipelines
or energy grids, are ubi\-quitous in highly technological
societies. Since the well functioning of these societies depends heavily
on transportation networks,
large-scale (cascading) failure are in particular threatening. 
Previous work on cascading failures in networks have often 
analyzed the occurrence of past failures \cite{chen2005,bao2009b} or
 studied phase transitions,
as a function of some network parameter, from a resilient
to a failure phase \cite{deArcangelis1985,crucitti2004,dobson2007,bao2009}. 
In this work we are concerned always with the
case of a design in such a way that a failure is prevented, i.e., 
resilient networks,
the typical task of an engineer. In particular, we are interested
on how to make a network fail-safe against the failure of one link
by including enough, but
as small as possible, \emph{backup capacity}, such that a cascading failure
is prevented (called ``$N-1$ criterion'' for power transmission).
 To gain a fundamental understanding of the
problem, no real-world networks are studied here. Instead,
this study is performed for three different network ensembles, two of
them  are highly relevant for transport processes in spatial settings, 
another simple model is included for comparison.
Here, the behavior over the range of (almost) each \emph{complete} 
ensemble is
addressed, this means in particular the properties of typical as well as
extremely resilient  and extremely weak networks are investigated. Namely, the
distribution of backup capacities is obtained for almost the complete
support, for backup capacities which appear about  $10^{-80}$ less likely than
in typical networks. This requires to apply 
special but simple numerical \emph{importance-sampling techniques},
as explained in section \ref{sec:method}, to obtain the
\emph{large-deviation} properties of the networks.

For many problems in science and in statistics, the large-deviation
properties play an important role \cite{denHollander2000,dembo2010}.
Only for few cases analytical results can be obtained.
Thus, most problems have to be studied by numerical simulations 
\cite{practical_guide2009}, in particular by Monte Carlo (MC) techniques
\cite{newman1999,landau2000}. Classically, MC simulations have
been applied to ensembles of random systems in the following way:
For a finite number of independently drawn  instances from the ensemble, 
arbitrary properties of these instances have been
calculated using MC simulations. 
Later, it has been noticed  that MC simulations can be
used via introducing an artificial
``temperature'' to 
sample the random ensemble such that the
large-deviation properties of the (almost) full ensemble
can be obtained \cite{align2002}. The results are 
re-weighted in a way that the results for 
the original quenched ensemble are obtained. In this way, the
large-deviation properties of the distribution of alignment
scores for protein comparison were studied 
\cite{align2002,align_long2007,newberg2008}, 
which is of importance to calculate
the significance of results of protein-data-base queries \cite{durbin2006}.

Motivated by these results, similar approaches have been applied to other
problems like the distribution of the number of components of
Erd\H{o}s-R\'enyi (ER) random networks \cite{rare-graphs2004},
the size of the largest components of ER random networks and of
two-dimensional grids \cite{largest-2011},
the partition function of Potts models \cite{partition2005},
the distribution of ground-state energies of spin glasses
\cite{pe_sk2006} and of directed polymers in random media \cite{monthus2006},
the distribution of Lee–-Yang zeros for spin glasses \cite{matsuda2008},
the distribution of success probabilities of error-correcting codes
\cite{iba2008}, the distribution of free energies of RNA secondary
structures \cite{rnaFreeDistr2010}, and some large-deviation properties
of random matrices \cite{driscoll2007,saito2010}.

To the knowledge of the author, no corresponding study has been 
performed to obtain the large-deviation properties of transport networks,
in particular of failure-resilient networks. 
Here, the large-deviation approach is applied to a simple yet often used 
transport model on three standard random network ensembles. 
Thus, this work serves in particular as a proof of principle that 
measuring large-deviation properties of transport networks
is possible and allows one to obtain useful insight. This 
shows that similar approaches
should be applicable to more complex transport networks, e.g., dynamic networks
of oscillators as used to study energy grids \cite{filatrella2008,rohden2012}.

The remainder
of the paper is organized as follows. First, in Sec.\ \ref{sec:ensembles},
the different network ensembles
under scrutiny are presented. Then, the backup capacity is introduced,
which is used to describe how resilient a network is. In Sec.\
\ref{sec:tests} a couple of test simulations are presented,
which explore the concept of the backup capacity. In the
following section, the large-deviation approach is presented.
Within the main Sec.\ \ref{sec:results}, all results are given. The paper
is closed by a summary and a discussion.

%
%
\section{Network Ensembles} 
\label{sec:ensembles}

This work is about the resilience of network ensembles, since such
ensembles are used often in theoretical studies about various network
properties. This type of approach is different from the question
how, e.g., the most resilient network for a given set of nodes and real-space
positions
looks like. For such a setup the notion of an ensemble makes less sense
and is thus not covered in this work. Also no existing networks are studied
empirically here since that would be beyond the scope of the work.
 Nevertheless, the present approach allows to gain insights
about the behavior of typical and atypical network instances, leading to 
general design principles for resilient networks.

\begin{figure}[t!]
  \centering
  \includegraphics[clip,width=0.3\textwidth]{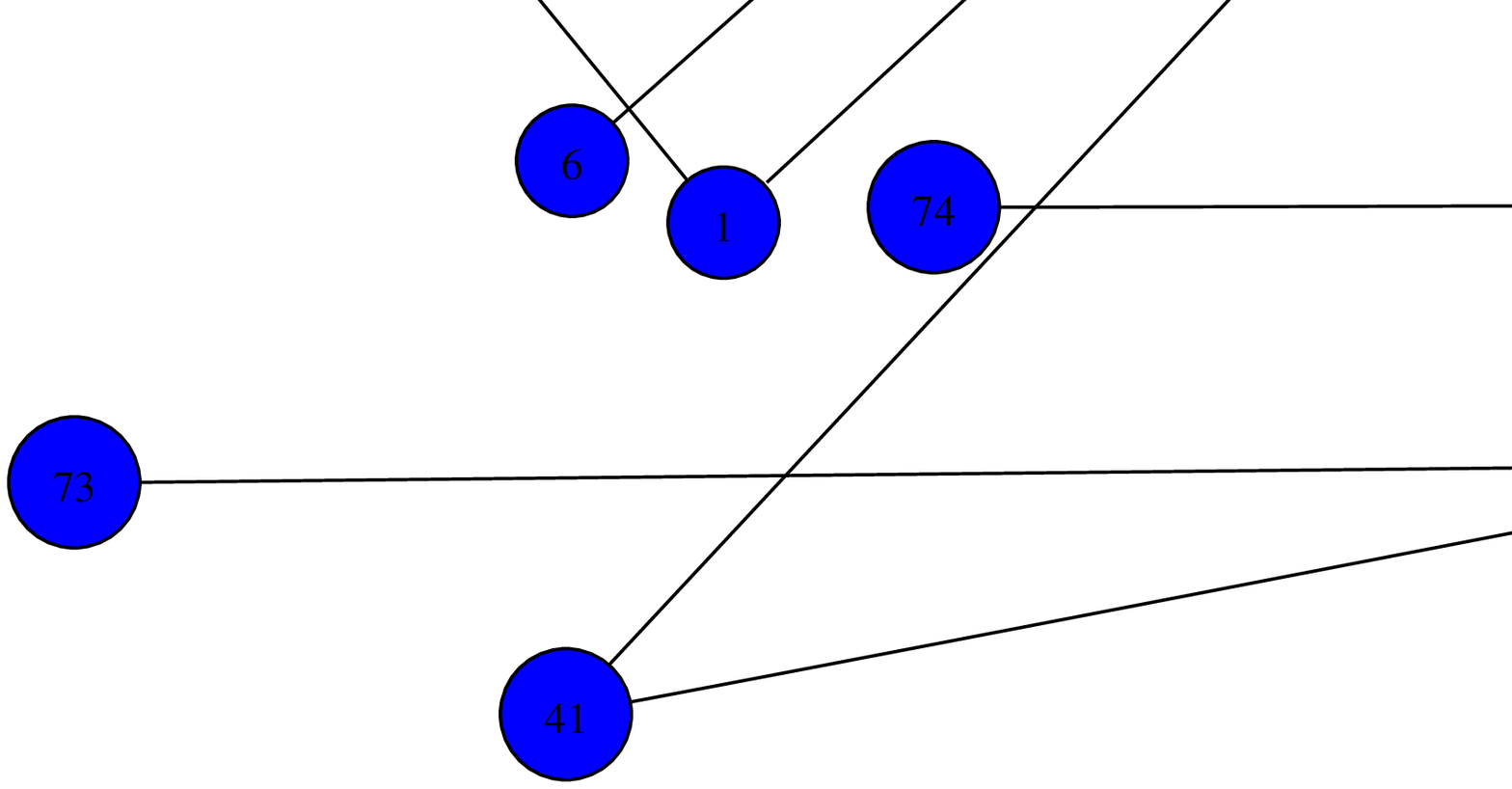}
  \includegraphics[clip,width=0.3\textwidth]{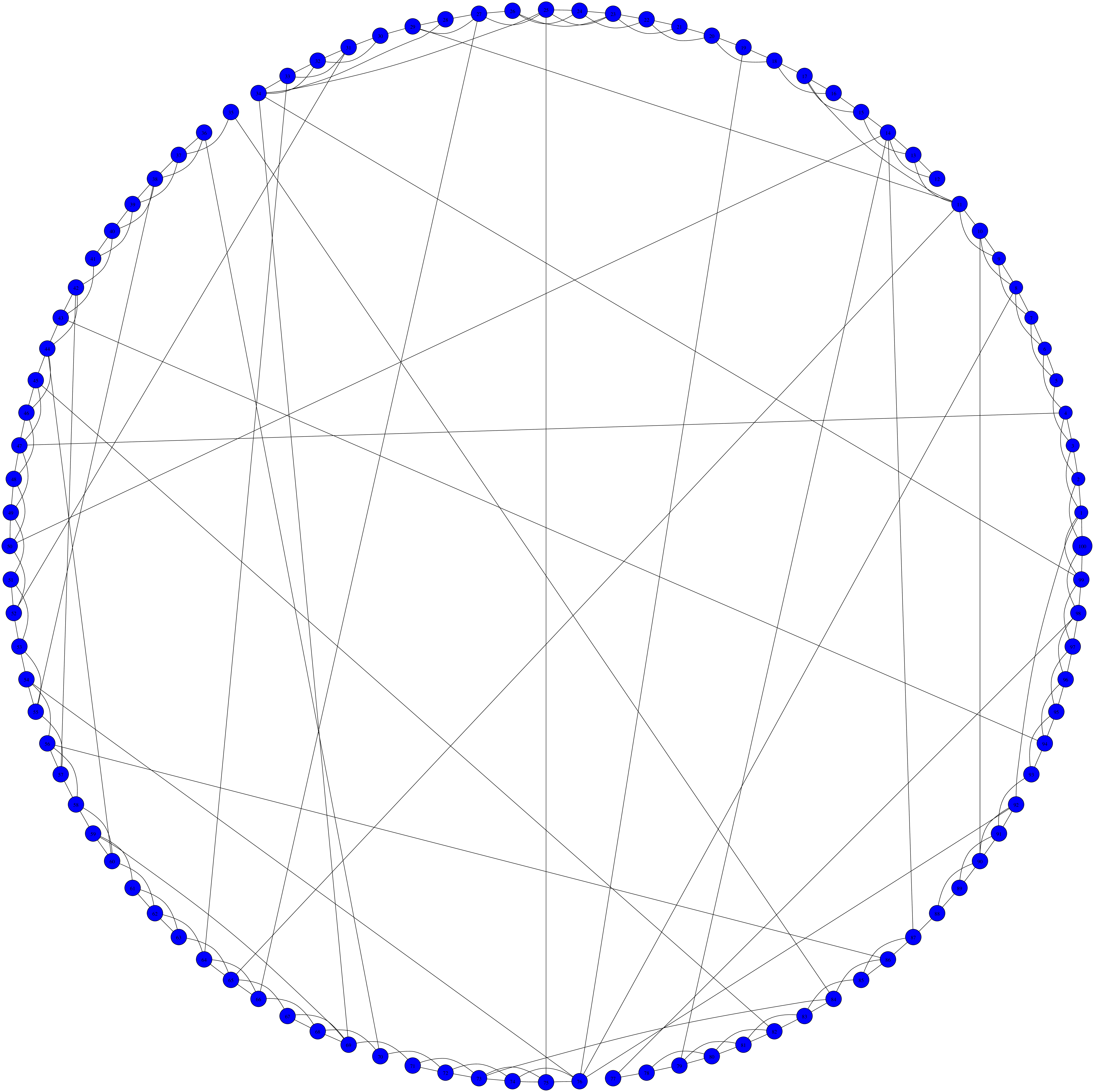}
  \includegraphics[clip,width=0.3\textwidth]{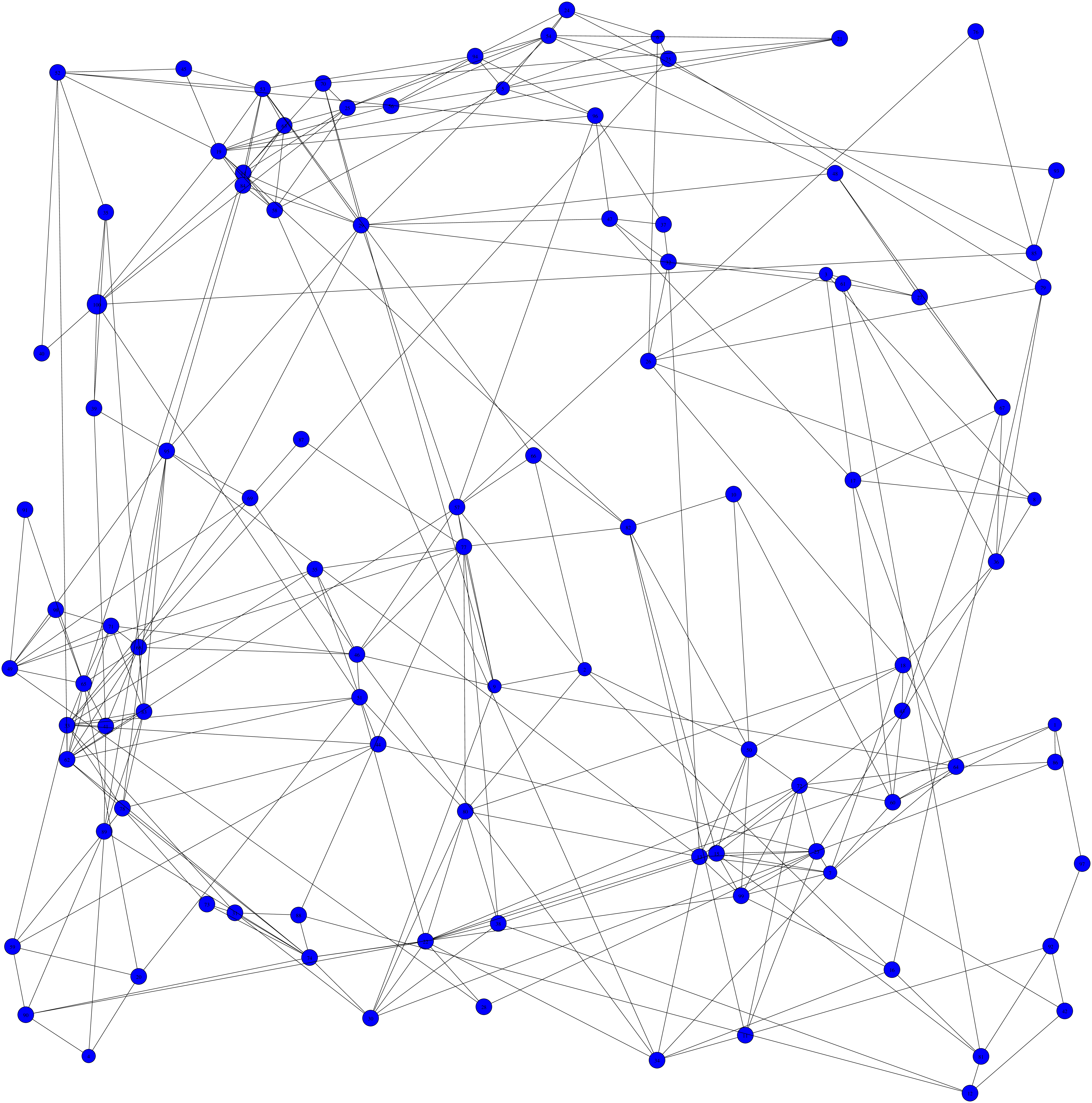}
  \caption{
    \label{fig:samples}
Three different ensembles are treated. For each ensemble, 
a sample network with $N=100$ nodes is shown:
Erd\H{o}s-R\'enyi random networks (top, with randomly placed
nodes), small-world networks  (middle)
and two-dimensional spatial networks (bottom).
}
\end{figure}

The most simple type of random networks
is the Erd\H{os}-R\'enyi (ER) 
network  ensemble \cite{erdoes1960}. It makes no assumption about the
topology of the network, i.e., in particular it exhibits no spacial structure,
see Fig.\ \ref{fig:samples} (top). Thus, it
 is ideally suited for comparison with more
complex network ensembles as well as single network instances 
to see what effect the structures of these more complex
networks have regarding their behavior.
To create an
ER random network, one starts with an empty network of $N$ nodes.
For each pair $i,j$ of nodes, with some given probability $p_{ij}^{\rm ER}$
 the link $\{i,j\}$ is added to the network. Here, 
\begin{equation}
p_{ij}^{\rm ER}=c/(N-1)
\end{equation}
is chosen such that, on average, each node
has $c$ connections. Nevertheless, all possible networks have a nonzero
(although sometimes quite small) probability, even the complete (fully connected) 
as well as the empty network.
Since ER random networks have a minimum
amount of structure, 
they often serve as a suitable null model
when comparing to other ensembles of random networks.

Next, a  widely studied model of networks is considered
here, the \emph{small-world} (SW) ensemble 
\cite{watts1998,amaral2000,barrat2001}. This ensemble was found, e.g., to represent
the U.S. power grid well \cite{watts1998,amaral2000,motter2002}
and was used for modelling other transport networks as well \cite{bao2009}.
For this model, in a first step
$N$ nodes are distributed on a ring and connected
with their direct and second-nearest neighbors. 
Thus, each node has four links.
Next, for each existing
link, with probability $p$ (here $p=0.1$) it is disconnected at one terminal
node and reconnected with a completely randomly chosen node, hence
keeping the average number of links per node. 
Thus, for $p\to 1$ the SW networks
become more similar to a modified ER
 ensemble where the average number of neighbors is $c=4$ and where the
actual number of links does not fluctuate. For $p$ small, this results
in a mixture of many local short-range and few long-range links,
 see Fig.\ \ref{fig:samples} (middle), which is a key characteristics of
 many existing real-world 
networks. Note that for easy comparison between different network models,
$c=4$ was chosen here for the ER ensemble, such that all network ensembles
have the same average number of neighbors.

Since many existing transport networks are embedded on a two-dimensional
(earth) surface, also a  \emph{spatial} (two-dimensional) 
model \cite{barthelemy2011} 
is included in the present study, which exhibits even more
spatial structure than the SW ensemble. Here, $N$ nodes are distributed
randomly with uniform weight in a $[0,1]^2$-plane. Afterward,
for each pair $i,j$ of nodes, with probability 
\begin{equation}
p_{ij}^{\rm sp}=f\cdot (1+\sqrt{N\pi}d_{ij}/\alpha)^{-\alpha}
\end{equation} 
the link $\{i,j\}$ is added to the network, where $d_{ij}$ is the 
Euclidean distance between nodes $i,j$. 
A sample network is shown in Fig.\ \ref{fig:samples} (bottom).
Here, values $f=0.95$ and $\alpha=3$
are chosen, which results also in an average number of neighbors
close to $4$.  Note that several variants
of spatial networks exist in the literature.
Although the model appears to be in particular useful
for surface-embedded transportation networks,
it is so far less established than the SW model, 
 so the many results  for the SW
model are included here as well.

\section{Resilience}
\label{sec:resilience}

The quantity to describe the resilience of a network against
a failure leading to cascading failures is based on a rather simple 
(i.e., not time-dependent) yet
established quantity to measure loads in transport networks
\cite{motter2002,motter2004,bao2009}.
The loads are given by the assumption that for each pair $i,j$
of nodes, a unit one of some quantity has to be transported
between $i,j$. This requires the network to be connected, i.e., to consist of a
single connected component. For the above random ensembles it means that they
are restricted to the subset of connected network instances.
For the SW model and the spatial model 
basically all network instances are connected using the given parameters,
while for the ER model typically only a small fraction
of networks is connected (37\% for $N=50$, 15\% for $N=100$, 2.5\% for $N=200$
and 0.005\% for $N=400$). Note that the sampling used here, see
Sec.\ \ref{sec:method}, ensures that only connected networks are sampled.

For the transport between any pair $i,j$ of nodes the shortest path is chosen
(if several shortest paths exist, the transportation
is divided equally among the different paths). This is performed
for all pairs of nodes, which are connected in the network. The actual
load  
$l_{i,j}$ for an link $e=\{i,j\}$ is the total number of (possibly sums
of fractional) shortest 
paths which run through the link. Hence, the load is equal to the
well known \emph{edge-betweenness}, which can be calculated 
conveniently using a fast algorithm \cite{newman2001shortest-path}.

Now, the \emph{backup capacity} $c_{\rm b}$ is defined. For this purpose,
the link 
$ e_{\max}={\rm argmax}_{\{i,j\}}\, l_{i,j}$
exhibiting the highest load is removed from
the network. Next, all loads are recalculated, resulting in load
values  $\{\tilde l_{i,j}\}$
of the modified network. The \emph{backup capacity} is defined as
the highest increment in the load, i.e.,
\begin{equation}
c_{\rm b} = \max_{\{i,j\}} (\tilde l_{i,j} - l_{i,j})\,.
\end{equation}
If the network is disconnected by the removal of $ e_{\max}$,
$c_{\rm b}=\infty$ is chosen, i.e., such networks are actually
disregarded as well.
The (or one) link which exhibits the maxi\-mum increase in load is
denoted here by $e_{\rm incr-max}$.
Note that in some link the load will actually decrease, but that
is no problem for the definition.
Hence, the backup capacity represents a rough and rather safe
estimate of how much the capacities have to be chosen above
the actual load values to make the transportation network resilient against
the failure of one link.

\section{Test simulations}
\label{sec:tests}

\begin{figure}[t!]
  \centering
  \includegraphics[clip,width=0.49\textwidth]{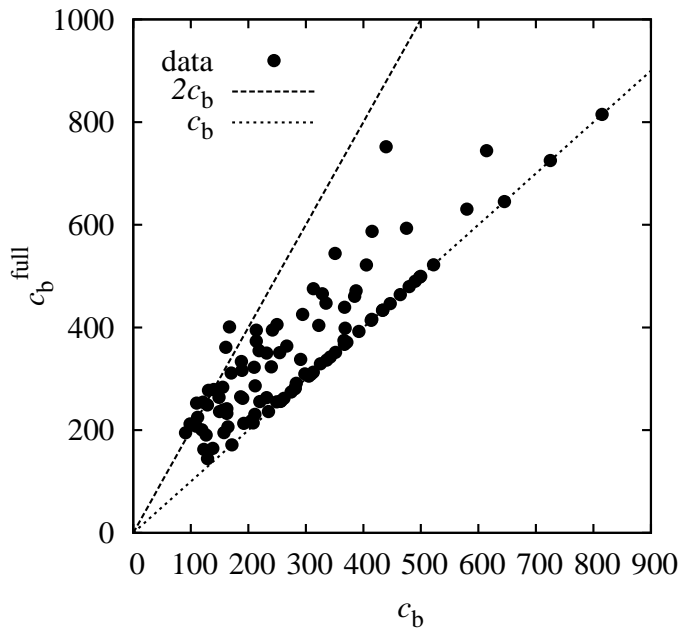}
  \caption{Scatter plot of the full  backup capacity 
$c_{\rm b}^{\rm full}$, which is optimized over the removal
of every link, one at a time, as a function of
actually used backup capacity $c_{\rm b}$ (where only the link with the
largest load is removed), for small-world networks with $N=100$ and $p=0.1$.
    \label{fig:resilience_full}
}
\end{figure}

To verify whether concentrating on the removal of the link with the
largest load (leading to backup capacity $c_{\rm b}$) 
is justified, simulations were performed such that
for each network the necessary backup capacity was also maximized over
the single-link removal of \emph{all links}, 
resulting in a maxi\-mum backup
capacity  $c_{\rm b}^{\rm full}$. 
Clearly, $c_{\rm b}^{\rm full}\ge c_{\rm b}$ holds for any network.
In Fig.\ \ref{fig:resilience_full}, 
results for small-world networks with $N=100$ nodes
and rewiring probability $p=0.1$ are displayed, 
the results for other network ensembles look
simi\-lar. As visible, there
is a strong correlation between the full backup capacity 
$c_{\rm b}^{\rm full}$ and the actually used backup capacity $c_{\rm b}$.
A linear correlation coefficient $0.85$ was found.
In fact, for about 1/3 of all networks, both are exactly equal
and for more than 90\% $c_{\rm b}^{\rm full}\le 2 c_{\rm b}$ holds.
Hence, due to the strong correlation, when optimizing the network topology
with respect $c_{\rm b}$ one will also obtain very efficient networks
with respect to $c_{\rm b}^{\rm full}$.

\begin{figure}[t!]
  \centering
  \includegraphics[clip,width=0.49\textwidth]{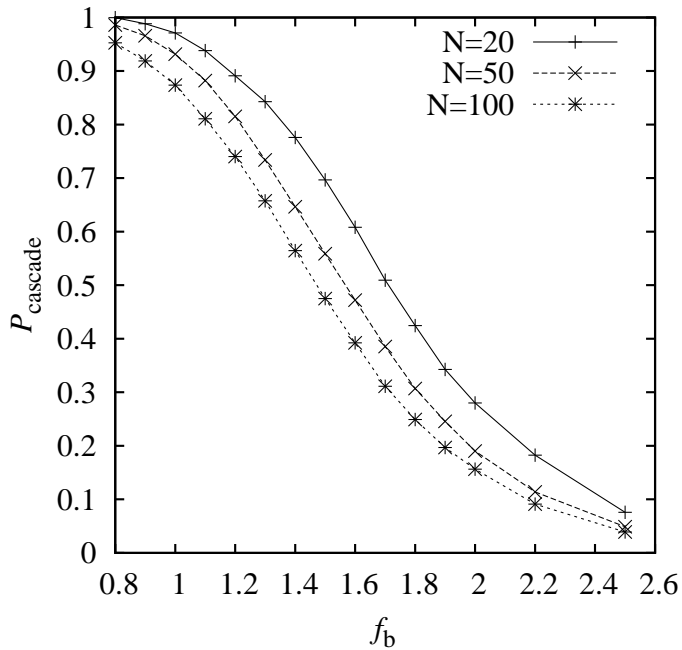}
  \caption{Probability $P_{\rm cascade}$ of a cascade which disconnects
the network as a function of the fraction $f_{\rm b}$ of the backup
capacity added to the link capacities.
 For small-world networks with different sizes $N=20$, $N=50$, 
$N=100$ and $p=0.1$.
    \label{fig:cascade}
}
\end{figure}

Thus, for best efficiency, for the main  simulations 
the above quantity $c_{\rm b}$ was evaluated, i.e.
 only one link was removed and the
load recalculated. Nevertheless, in principle one is interested
in global events, i.e., in \emph{cascading failures} \cite{crucitti2004}. 
The basic assumption is
that when the backup capa\-city is not sufficient, a cascading failure
will be triggered frequently. This assumption was also checked explicitly
in this work for the sample setup of SW networks ($p=0.1$): After
the calculation of the backup capacity $c_{\rm b}$ the capacity of every
link was increased by the amount $f_b c_{\rm b}$, where $f_b$ is
a factor in the range $[0,2.5]$, except for the link $e_{\rm incr-max}$ 
 which exhibits the
largest load increase after the initial removal. For this link
the capacity is increased only by $c_{\rm p}-\epsilon$, where $\epsilon>0$. 
Thus, this link will for sure fail
next with the recalculated load distribution since the
capacity is not sufficient. Thus it has to be
removed as well and the load distribution has to be recalculated again. Now
even more links may fail. This process is repeated, until no more links fail,
i.e., the network is able to redistribute the load, or until 
a cascade of failures result in the
network breaking apart, i.e. a complete breakdown. 
In Fig.\ \ref{fig:cascade}, the proba\-bili\-ty
$P_{\rm cascade}$ of a cascading failure leading to a network breakdown
is shown as a function of $f_b$. Clearly, for $f_b<1$ almost
all failures trigger a network breakdown. 
 On the other hand, when increasing $f_b$ cascading failures
become less likely. Hence, it is justified to calculate just the
backup capacity $c_{\rm b}$, involving only two calculations
of the load distribution, to learn about the resilience of the
network against an event leading to a cascading failure.

%
%
\section{Simulation and reweighting method} 
\label{sec:method}

To determine the distribution $P(c_{\rm b})$ for any measurable quantity $c_{\rm b}$,
here denoting the backup capacity  of a network, \emph{simple sampling} is
straightforward: One generates a certain number $K$ of network samples
and obtains $c_{\rm b}(G)$ for each sample $G$.
This means each network $G$ will appear with its natural ensemble 
probability $Q(G)$.
The probabili\-ty to measure a value of $c_{\rm b}$ is given by
\begin{equation}
P(c_{\rm b}) = \sum_{G} Q(G)\delta_{c_{\rm b}(G),c_{\rm b}} \label{eq:PS}
\end{equation}
Therefore, by calculating a histogram of the values for $c_{\rm b}$, a good
estimation for $P(c_{\rm b})$ is obtained.
Nevertheless, $P(c_{\rm b})$ can only be measured in a regime where $P(c_{\rm b})$
is relatively large, about $P(c_{\rm b})>1/K$. Unfortunately, the distribution
decreases for many systems exponentially fast in the system size $N$ when moving
away from its typical (peak) value.
This means, even for moderate system
sizes $N$, the distribution will be unknown on almost its complete support.

To estimate $P(c_{\rm b})$ for a much larger range,  even possibly on the 
full support of $P(c_{\rm b})$, where probabilities smaller
than $10^{-10}$ may appear,
a different approach is used \cite{align2002}. 
For self-containedness, the method is outlined subsequently.
The basic idea is to use an additional
Boltzmann factor $\exp(-c_{\rm b}(G)/T)$, $T$ being a ``temperature''
parameter,  in the following manner:
A Markov-chain MC simulation \cite{newman1999,landau2000}
is performed, where in each step
$t$ from the current network $G(t)$ a candidate network $G^*$ is created:
A node $i$ of the current network is selected randomly, 
with uniform weight $1/N$,
and all adjacent links are deleted. For all feasible links $\{i,j\}$, the link
is added with a weight corresponding to the natural weight $Q(G)$, e.g.,
with probability $c/(N-1)$ for ER random networks. For SW and spatial
networks it is done correspondingly, see Sec.\ \ref{sec:ensembles}. 
Next, it is checked whether network $G^*$ is connected, i.e., consists of one single
connected component, because only for a connected network the backup capa\-city
is defined. If $G^*$ is not connected, it is rejected, i.e. $G(t+1)=G(t)$.
Note that it has to be made sure that the initial network is connected. This
was achieved by gene\-rating candidates for the initial network until a connected
instance was found.
 
If the candidate network is connected,
 the backup capacity $c_{\rm b}(G^*)$ is calculated. Finally,
the candidate network is \emph{accepted}, ($G(t+1)=G^*$) 
with the Metropolis proba\-bility
\begin{equation}
p_{\rm Met} = \min\left\{1,e^{-[c_{\rm b}(G^*)-c_{\rm b}(G(t))]/T}\right\}\,.
\end{equation}
Otherwise the candidate network is also rejected ($G(t+1)=G(t)$).  By construction,
the algorithm fulfills detailed balance. Clearly the algorithm is also
ergodic, since within $N$ steps, each possible network may be constructed, in
principle. Thus,
in the limit of infinite long Markov chains,
the distribution of networks will follow the probability
\begin{equation}
q_T(G) = \frac{1}{Z(T)} Q(G)e^{-c_{\rm b}(G)/T}\,, \label{eq:qT}
\end{equation}
where $Z(T)$ is the a priori unknown normalization factor.

The distribution for $c_{\rm b}$ at temperature $T$ is given by
\begin{eqnarray}
P_T(c_{\rm b}) & = &\sum_{G} q_T(G) \delta_{c_{\rm b}(G),c_{\rm b}} \nonumber\\
       & \stackrel{(\ref{eq:qT})}{=} & 
         \frac{1}{Z(T)}\sum_{G} Q(G)e^{-c_{\rm b}(G)/T} \delta_{c_{\rm b}(G),c_{\rm b}} \nonumber \\
       & = & \frac{e^{-c_{\rm b}/T}}{Z(T)} \sum_{G} Q(G) \delta_{c_{\rm b}(G),c_{\rm b}}
               \nonumber \\
       &  \stackrel{(\ref{eq:PS})}{=} & 
           \frac{e^{-c_{\rm b}/T}}{Z(T)} P(c_{\rm b}) \nonumber\\
\Rightarrow \quad P(c_{\rm b}) & = & e^{c_{\rm b}/T} Z(T) P_T(c_{\rm b}) \label{eq:rescaling}
\end{eqnarray}
Hence, the target distribution $P(c_{\rm b})$ can be estimated, up to a normalization
constant $Z(T)$, from sampling at finite temperature $T$. For each
temperature, a specific range of the distribution $P(c_{\rm b})$ will be sampled:
Using a positive temperature allows to sample the region of
a distribution left to its peak (values
smaller than the typical value), 
while negative temperatures are used to access the right tail.
Temperatures of large absolute value will cause a sampling of the 
distribution close to its typical value, while temperatures of small 
absolute value
are used to access the tails of the distribution. Hence, by choosing a
suitable set of temperatures, $P(c_{\rm b})$ can be measured over a large
range, possibly on its full support.

The normalization constants $Z(T)$ can easily be obtained by including
a histogram obtained from simple sampling, which corresponds
to temperature $T=\pm\infty$, which means $Z\approx 1$ 
(within numerical accuracy). Using suitably chosen
temperatures $T_{+1}$, $T_{-1}$, one measures histograms which overlap 
with the simple sampling histogram on its left and right border,
respectively. Then the corresponding normalization
constants $Z(T_{\pm 1})$ can be obtained by the requirement that
after rescaling the histograms according to 
(\ref{eq:rescaling}), they must agree
in the overlapping regions with the simple sampling histogram within
error bars. This means, the histograms are ``glued'' together,
similar to the multi-histogram approach of Ferrenberg and Swendsen 
\cite{ferrenberg1989}. In the same
manner, the range of covered 
$c_{\rm b}$ values can be extended iteratively to the left and to
the right by choosing additional suitable temperatures 
$T_{\pm 2}, T_{\pm 3}, \ldots$ and gluing the 
resulting histograms one to the other. A pedagogical explanation
 and examples of 
this procedure can be found in Ref.\ \cite{align_book}.

In order to obtain the correct result, the MC simulations must be
equilibrated. The equilibration of the simulation can be monitored by starting
with two different initial networks, respectively: 
\begin{itemize}
\item First an unbiased 
random network is taken, which means that the measure of interest is
close to its typical value.

\item Second, one uses a very atypical network, e.g., a fully connected
network. 
\end{itemize}

In any case,
for the two different initial conditions,
the evolution of $c_{\rm b}(G(t))$  will approach from two different
extremes, which allows for a simple equilibration test:
equilibration is achieved if
the measured values of $c_{\rm b}$ agree within the range of fluctuations.
 For the simulations performed in this work,
equilibration was achieved always within 200 Monte Carlo sweeps (i.e., $200*N$
Monte Carlo steps).

%
%
\section{Results}
\label{sec:results}

Simulations where performed for ER, SW and spatial networks. For
each type, several number of nodes were considered, 
to study finite-size effects. The evaluation of the backup
capacity is rather involved, compared, e.g., to past large-deviation
studies of the largest component of networks \cite{largest-2011}. 
Thus, the largest networks
under scrutiny here exhibit $N=400$ nodes. Nevertheless, many 
existing transportation networks are of similar size.

\begin{figure}[t!]
  \centering
  \includegraphics[clip,width=0.49\textwidth]{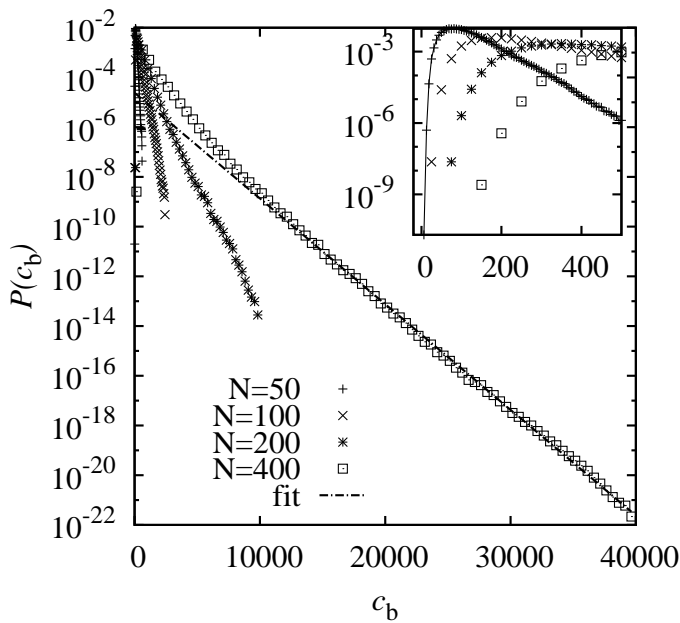}
  \caption{Probability distribution $P(c_{\rm b})$ for the backup capacity
$c_{\rm b}$, 
 for small-world networks ($p=0.1$) with different sizes $N=50$, 
$100$, $200$ and $400$.  Standard error bars are at most of 
order of symbol size. The line shows a fit of the tail ($c_{\rm b}\ge 20000$)
of the  $N=400$ data
to an exponential ($a\exp({-c_{\rm b}/\tilde c})$, $a=2.4(1)\times 10^{-5}$, 
$\tilde c=1022(1)$). The inset enlarges the region near
$c_{\rm b}=0$, the line being a guide to the eyes only.
    \label{fig:distr}
}
\end{figure}

Figure \ref{fig:distr} shows the distribution of the backup capacity for
almost the full support for the ensemble of SW networks. Note that
probabilities as small as $10^{-22}$ are easily obtained which are
clearly out of reach using conventional simulation techniques.
Typical, very reliable and very unreliable networks are accessible
using the large-deviation approach.
Typical networks, near the peak of the distributions, exhibit a rather
small backup capacity. Very unreliable networks, where the backup capacity has
to be large to prevent cascading failures, are very rare and located
in the tails of the distributions to the right. The tails of the distribution
follow exponentials very well, as a fit to $a\exp({-c_{\rm b}/\tilde c})$
to the tail of the data for the largest networks revealed,
resulting in $a=2.4(1)\times 10^{-5}$, 
$\tilde c=1022(1)$. Below a more detailed analysis via an extrapolation of
the large-deviation rate function is given, supporting that the limiting
distribution is exponential.

\begin{figure}[t!]
  \centering
  \includegraphics[clip,width=0.39\textwidth]{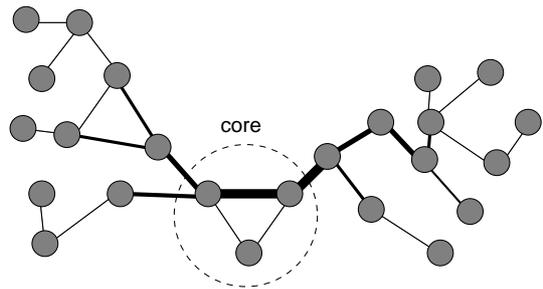}
  \caption{Network with highest backup capacity. Two large $O(N)$
subnetworks are connected trough a small core network (a triangle, here). 
The links
with the highest loads are shown in bold. Removing the highest load
from the core part results in an increase of the load through the
other core links by an amount $N^2/4$.
    \label{fig:worst_case}
}
\end{figure}

An inspection of the networks in the far tail of the distribution 
showed that the most unreliable networks have a quite
special structure. They consist of a small core of connected nodes,
e.g., a triangle of nodes in the simplest case,
see Fig.\ \ref{fig:worst_case}.  All remaining ($N-3$)
nodes are connected, directly or indirectly, to one of two of these
core nodes,  roughly partitioned equally into two sets, i.e. about $N/2$
per set of nodes.
 This means, a large number of $(N/2)^2$ shortest-path connections
runs from one set through a single link of the core to
the other set. This single link exhibits the highest load,
while the other core links are not used much. By removing this 
extreme-load link,
the load is redistributed completely in the core, hence, increased
by an amount of $\sim N^2/4$. Clearly, such networks are very
rare by chance and rather special. 

On the other hand, there are also networks which have even a 
much more resilient structure than typical
networks, since they require only a rather small backup capacity. They
are located near the origin of the distribution and are also rather
rare (probability $< 10^{-9}$ for the largest case considered here). Below
we will identify some structural network properties that make 
it very resilient.

The inset of Fig.\ \ref{fig:distr} shows also that
with increasing network size, the typical backup capacity, i.e.,
 the location of the peak,  grows.
A more detailed study, also invol\-ving larger sizes which were studied by
simple sampling, exhibits that the growth is linear (not shown here).

\begin{figure}[t!]
  \centering
  \includegraphics[clip,width=0.49\textwidth]{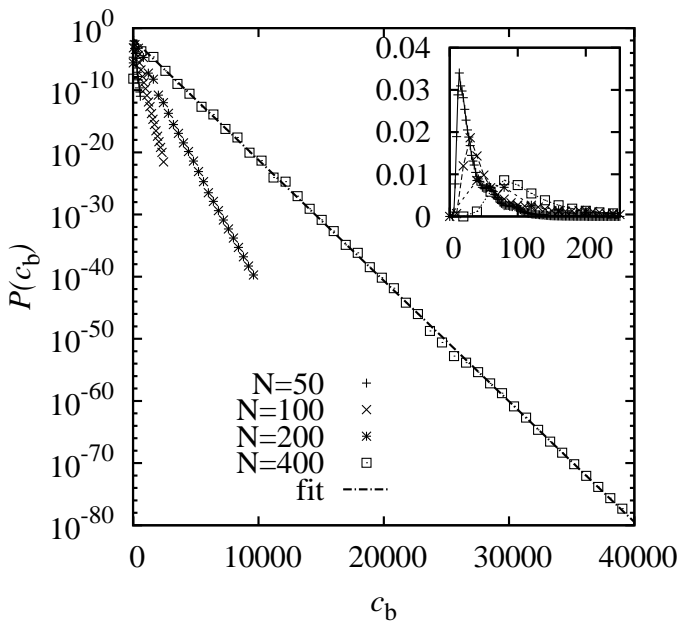}
  \caption{Probability distribution $P(c_{\rm b})$ for the backup capacity
$c_{\rm b}$, 
 for Erd\H{os}-R\'enyi networks ($c=4$) with different sizes $N=50$, 
$100$, $200$ and $400$.  Standard error bars are at most of 
order of symbol size. The line shows a fit over almost the full support
($c_{\rm b}\ge 5000$)  of the $N=400$ data
to an exponential ($a\exp({-c_{\rm b}/\tilde c})$, $a=0.015(2)$, 
$\tilde c=223(1)$). The inset enlarges the region near
$c_{\rm b}=0$, lines being guides to the eyes only.
    \label{fig:distr:er}
}
\end{figure}

For comparison with the most simple network model, Erd\H{os}-R\'enyi (ER) 
random networks \cite{erdoes1960}, the corresponding results are shown
in Fig.\ \ref{fig:distr:er}. The peak of the distribution again 
moves (linearly) to the right but is located relatively much 
further left compared to the SW case.
Again an exponential fits the data well, now for almost over the full support.

\begin{figure}[t!]
  \centering
  \includegraphics[clip,width=0.49\textwidth]{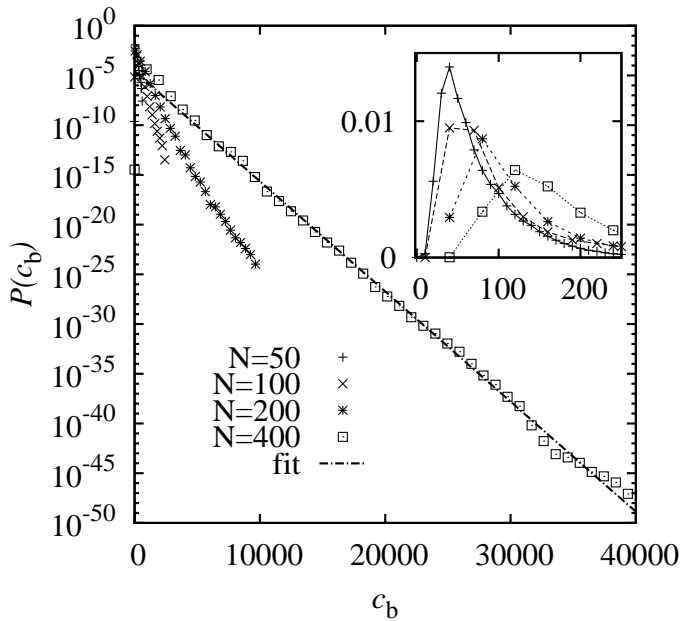}
  \caption{Probability distribution $P(c_{\rm b})$ for the 
backup capacity, for spatial random networks 
($\alpha=3$, $f=0.95$) with different sizes $N=50$, $N=100$, $N=200$ 
and $N=400$. Standard error bars are at most of order of symbol size.
The line shows a fit over most of the support
($c_{\rm b}\ge 10000$)  of the 
$N=400$ data
to an exponential ($a\exp({-c_{\rm b}/\tilde c})$, $a=2.7(4)\times 10^{-5}$, 
$\tilde c=392(1)$). The inset enlarges the region near $c_{\rm b}=0$, 
lines being guides to the eyes only.
    \label{fig:distr:spatial}
}
\end{figure}

For comparison with the realistic spatial network model, 
the corresponding results are shown
in Fig.\ \ref{fig:distr:spatial}. Typically, these networks are more resilient 
(smaller value of $c_{\rm b}$) than the SW networks, but less resilient 
than the ER model. Again,
most of the distribution can be well fitted by an exponential.

Comparing the insets of Fig.\ \ref{fig:distr}, \ref{fig:distr:er} and
\ref{fig:distr:spatial}, one observes that both ER random networks as well
spatial random networks typically require a lower backup capacity
compared to the SW model. 
Typical values of the backup capacity $c_{\rm b}$ for the ER model are 
located at small values ($c_{\rm b}\approx 40$ for $N=400$). The 
spatial networks
are typically almost as resilient ($c_{\rm b}\approx 100$ for $N=400$). 
For the SW ensemble, typical networks need much larger backup capacity 
($c_{\rm b}\approx 500$ for $N=400$).
Correspondingly, large values of $c_{\rm b}$ are very unlikely for the
ER ensemble (a density of 
$10^{-80}$ at $c_{\rm b}=40000$ for $N=400$) and quite unlikely for 
the spatial networks (a density of  $10^{-50}$). 
For the SW model the density of $10^{-22}$ at rightmost tail is relatively 
larger. These quantitative differences in the large-deviation behavior 
 are also reflected by the constants obtained by fitting an exponential
to the tail of the distribution. The drop of the tail is strongest for the
ER model, followed by the spatial ensemble and finally by the SW networks.
Thus, the SW ensemble relatively favors less resilient networks compared
to the other two ensembles.
For the ER model this behavior is no
big surprise
because the ER model does not exhibit any spatial structure,
allowing for arbitrary network topologies in particular many 
long-range links (equivalent to
a high dimension of the system) which lead to a strong resilience.
In particular the network where all pairs of nodes are directly
connected (complete network) is contained in the ensemble, which
is not possible for the SW ensemble because the average number of neighbours
is fixed to 4.

On the other hand, the SW and the spatial model are both embedded
in a low-dimensional structure, which might indicate that both should need
larger backup capacities than the ER ensemble. Nevertheless, the
spatial ensemble seems to be more similar to the ER networks with respect
to the resilience. First, one should note that for the spatial networks 
in principle a complete network is possible (but even more unlikely 
than for ER random networks)
in contrast to the SW ensemble. Second, the SW model exhibits indeed 
some long-range links, which can be used to decrease to overall load, hence
the backup capacity. Nevertheless, by accident few of the long-range links
will be  very suitable for many of the shortest paths, acquiring
much of the load, while many other long-range links carry only a small load.
This ``channeling effect'' leads to a rather large backup capacity, 
thus a large load has to be rerouted after the failure. 
Opposite to this, for the spatial model, due to the 
distribution of the nodes in a two-dimensional plane, the total 
all-to-all traffic is distributed more over different paths, 
leading to a more uniform distribution
of the load in the plane, in turn requiring less backup capacity.

Thus,
using a true two-dimensional model, like the one applied here, appears from
the present results the most meaningful approach within this field.
This type of networks exhibit on the one hand spatial structure, as needed
for most real-world applications, one the other hand, the resilience is typically,
and also optimally, rather large. This is opposed to the SW model, which 
is often used to model power grids 
\cite{watts1998,amaral2000,motter2002}, but exhibits rather low resilience
and is not embedded in a two-dimensional plane.
 Clearly, by increasing the
fraction $p$ of randomized links, the SW model can be made more resilient,
but that will render it much more similar to the ER model (without fluctuating
number of links), but less
finite-dimensional, i.e., less realistic.

\begin{figure}[t!]
  \centering
  \includegraphics[clip,width=0.49\textwidth]{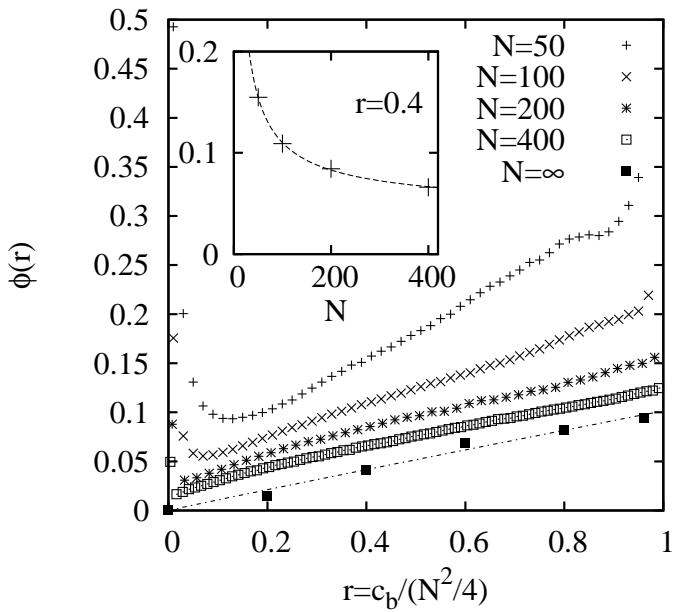}
  \caption{Rate function $\phi$ for the rescaled backup capacity
$r=c_{\rm b}/(N^2/4)$, 
 for small-world networks ($p=0.1$) with different sizes $N=50$, 
$N=100$, $N=200$, $N=400$ and for the extrapolation $N\to\infty$. 
The line represents a power law fit to the extrapolated values.
The
inset shows a sample extrapolation (using a power law plus
constant) for $r=0.4$.
    \label{fig:rate_fct_sw}
}
\end{figure}

Next, some information about the limiting distribution for large network
sizes is  obtained using the so-called \emph{rate function}
\begin{equation}
\phi = - \frac 1 N \log P(c_{\rm b})
\end{equation}
which is a standard quantity in large-deviation theory 
\cite{denHollander2000,touchette2009}. It displays the leading behavior under
the assumption that away from the typical instances, the distribution 
decays exponentially fast in the system size, i.e., $P(c_{\rm b})\sim e^{-N\phi}$.
In Fig.\ \ref{fig:rate_fct_sw} the rate function 
as a function of the rescaled backup capacity
$r=c_{\rm b}/(N^2/4)$ is shown for SW networks. This rescaling to 
$r$, motivated by the above finding about the most unresilient networks,
 ensures that the
maximum of the support for the rate function is close to $r=1$. 
This allows for a comparison
and extrapolation of the results for different sizes $N$.
Just from looking at the data, the rate function seems to approach a limiting
shape for larger networks.

To make this statement quantitatively precise, an extrapolation
to $N\to\infty$ was performed in the following way: For selected
fixed values of $r$, the rate function $\phi$ was considered
as a function of the system size $N$ and fitted to a power
law $\phi_r(N)=\phi_r^{\infty}+b_r N^{-c_r}$, which is a typical
finite-size behavior found in statistical mechanics models. The inset
of Fig.\ \ref{fig:rate_fct_sw} shows the SW data and the resulting
fit for the case $r=0.4$ (with $\phi_{0.4}^{\infty}=0.041(7)$,
$b_{0.4}=1.9(6)$ and $c_{0.4}=0.72(9)$). The resulting
extrapolated values  $\phi_r^{\infty}$ are also displayed in 
Fig.\ \ref{fig:rate_fct_sw} together with a fit to a power
law $\phi_r^{\infty}=\alpha r^\beta$ ($\alpha=0.102(5)$,$\beta=0.97(10)$,
i.e., close to a linear behavior), which is compatible with
an exponential distribution ($\beta=1$) for the backup capacity in the thermodynamic
limit. For the two other network types, the exponential nature of
the tails of the distributions is even more obvious from the data 
shown in Figs.\ \ref{fig:distr:er} and \ref{fig:distr:spatial} directly,
hence corresponding analyses of the rate function are omitted here.
The fact that the data can be so well described by the rate function in the
thermodynamic limit indicates that the problem studied here may be well
accessible using analytical large-deviation approaches, which often are based
on obtaining the rate function.

Finally, we want to understand the source of resilience in principle. 
Trivially, the higher the load in the most-loaded link, the more load has to be
redistributed when this link is removed, i.e., the larger the needed backup
capacity. More interesting it is to ask
which network structures lead to resilient networks, without looking 
at the actual load values. Here, selected results are shown for
the connection between the resilience and the number of links and, 
respectively, the diameter of a network.

\begin{figure}[t!]
  \centering
  \includegraphics[clip,width=0.49\textwidth]{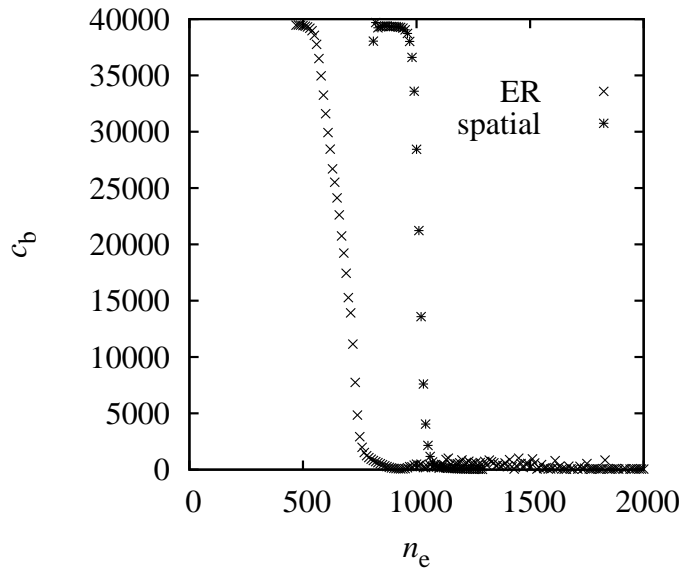}
  \caption{The average resilience $c_{\rm b}$ as a function of the
number $n_{\rm e}$ of links for ER and spatial networks of size $N=400$.
    \label{fig:correlation_edges}
}
\end{figure}

First, the relationship of the resilience to the number of links is 
investigated. For this purpose, the $N=400$ networks obtained during the 
simulations at different temperatures
$T$ were binned according to the number $n_{\rm e}$ of links. For the 
networks in each bin, the average backup capacity $c_{\rm b}$ was 
evaluated. The result is shown in Fig.\
\ref{fig:correlation_edges} for the ER and the spatial random networks (for the
SW ensemble, the number of edges does not vary). One sees that if only
few links are available, the backup capacity is very large, which 
is meaningful,
because having more links allows to distribute the load making a network more
resilient. Interestingly, a sharp drop as a function of $n_{\rm e}$ is visible,
looking like a phase transition. This drop appears for ER random networks at a
smaller number of edges, which is meaningful, because ER networks exhibit no 
constraints, thus one has more ``freedom'' to arrange the links such that
a high resilience can be obtained. Note that
 typical networks exhibit a
very small backup capacity compared the the ``high-backup capacity phase''
in the left part of Fig.\ \ref{fig:correlation_edges}. Thus, this
transition is not investigated more thoroughly here.

\begin{figure}[t!]
  \centering
  \includegraphics[clip,width=0.45\textwidth]{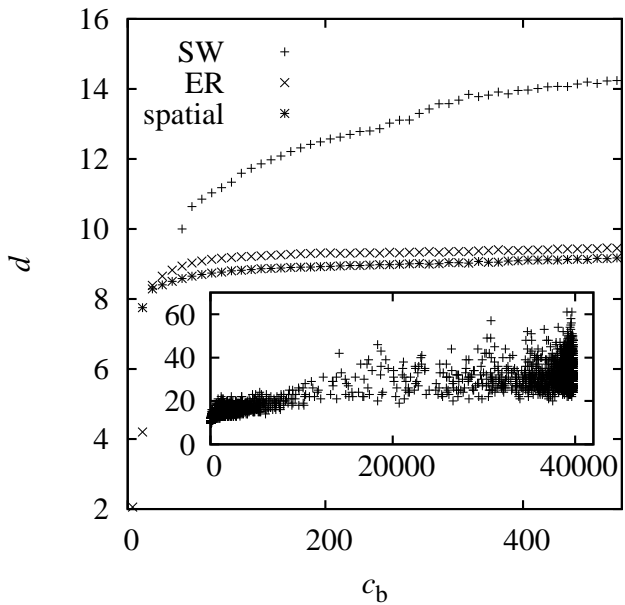}
  \caption{The average diameter $d$  as a function of the
resilience $c_{\rm b}$ for SW, ER and spatial networks of size $N=400$,
in the range of small value of $c_{\rm b}$.
The inset shows a scatter plot of the data for the entire range of 
backup capacities for the SW case.
    \label{fig:correlation_diameter}
}
\end{figure}

Thus, including more edges leads, not surprisingly, to a higher resilience. 
Note that examples exist, where adding more links sometimes also
decreases the stability of a network \cite{witthaut2012}.
Anyway,
for real networks, including more links leads almost always to larger 
costs (see also remarks below). Thus,
it would be interesting to see how the resilience correlates with other
topological measures of the network. For this purpose, a scatter plot of the
backup capacity versus the diameter $d$ of a network (i.e., the 
longest among all
shortest $i\leftrightarrow j$ paths) was recorded, see inset of Fig.\
\ref{fig:correlation_diameter} for the SW model. One can observe that large
backup capacities go along with large diameters. Note that for the SW model,
the differences in backup capacity can not originate from fluctuations of the
number of edges.
The positive relation between diameter and backup capacity can be 
seen even better in the main part of Fig.\ \ref{fig:correlation_diameter}, 
where a binning of the networks with respect to $c_{\rm b}$ was 
performed (shown here just for typical and very resilient networks,
i.e., small values of $c_{\rm b}$) 
and within each bin the average
diameter was evaluated. Again, the positive correlation between diameter and
backup capacity is visible, for all three network ensembles. In particular
very small backup capacities, i.e, the most resilient graphs (which are
not accessible using standard simple-sampling simulation approaches) are related
to extremely small diameters. This shows that the diameter is a key quantity when
considering and optimizing resilience of transport networks.
Note that again
ER ensemble and spatial networks are very similar behaving, although very
different in definition. For the ER ensemble, for the most resilient instances
obtained, an average diameter of $d\approx 2$ was measured, close to the value of $d=1$
of the complete network. Note that a complete network minus one edge has already
diameter two. This shows that actually the most resilient networks were sampled
during the simulations.

On the other hand, the SW ensemble exhibits for
the same backup capacity a larger average diameter. This may occur on the first sight
interesting since it allows for longer paths leading to the same resilience. On the
other hand, this effect is only visible for slightly higher backup capacities.
Therefore, in the region of extremely resilient networks, which are most interesting, the
SW ensemble does not contribute at all, because such resilient networks do simple
not exist there. Furthermore, extrapolating the SW data of $d$ by eye to small
values of $c_b$ leads to small values of the diameter, which simply cannot be
obtained in this ensemble.

Note finally that for real networks costs are an important issue.
In order to observe how the resilience scales with the costs
of a network, one would have to take into account the spatial length
of the links and built upon that the costs which are a function of the number
of links and their lengths. This is certainly beyond the
scope of the current work which has its focus on abstract but standard and widespread
network ensembles.  Nevertheless, the general results as
shown here will likely persist, in particular the strong correlation
with the diameter of a network, which should be minimized for given costs.

\section{Summary and outlook}
\label{sec:summary}

The resilience of simple models of transportation networks against
failures of highly-loaded links were studied here.
For the \emph{random networks}, three different ensembles were considered:
The Erd\H{o}s-R\'enyi ensemble is the
most simple model for random networks, exhibiting no spatial structure
at all, but serves well as a null model for comparison.
Small-world networks are also very simple but are used often to 
model real-world transportation
networks still rather well, like, e.g., energy grids.   Finally, spatial
networks are considered here, which are more sophisticated, 
but not well established.
They might serve in the future as standard models for
surface-based transportation.

To model the resilience against single-link failures (leading to
cascading failures) the \emph{backup capacity} is defined, which
describes the amount of additional capacity, which one has to
be included in the links to prevent a failure. The lower the backup
capacity is, the more resilient, i.e., the better, is the structure of the network.

Here, a \emph{large-deviation} approach was used to study
the distribution of the backup capacity.
Since the method allows to access a distribution (almost) on its complete
support, one can study the scaling behavior not only of the typical and
average but also of the best and the worst network instances.
Networks leading to very small probability
densities of the backup capacity  such as $10^{-80}$
could be generated and studied with the correct weight via introducing a
bias and reweighting the results for the analysis.

The main results are as follows: Trivially, by including more links,
a network can be made more resilient. More interestingly, for all types of  networks,
even for the SW ensemble with fixed number of links,
 the most-resilient
networks can be obtained by minimizing the diameter of the network.
The typical backup capacity, on the other hand, grows linearly
with the number of nodes in the network. In particular, spatial (two-dimensional)
networks appear most promising for future studies of resilience of models of
real-world transportation networks.

Furthermore, using the rate function
approach, the shape of the distribution could be extracted in the
thermodynamic limit, which is exponential. In particular, the
large-deviation property is fulfilled, which means that it appears
promising to use standard mathematical large-deviation techniques, e.g.,
generating functions, to study the distribution of the backup capacity
more rigorously. 

Hence, this study shows that the full range of transportation
networks ranging from the rare very resilient, over typical to the
exponential rare very susceptible networks can be studied numerically
using large-deviation techniques. Here, a rather simple and unspecific
transportation model, yet widely used in the literature,
 was used. Hence, it appears to be very promising
to apply similar approaches to more realistic and specific models
of transportation networks, e.g., time-dependent ac currents
based on Kuramoto oscillators to model energy grids 
\cite{filatrella2008,rohden2012,witthaut2012}.

\section*{Acknowledgements}

The author thanks Frank den Hollander for interesting discussions. The
author is grateful to Timo Dewenter for critically reading the manuscript.
Financial support was obtained via the
Lower Saxony research network ``Smart Nord'' which 
acknowledges the support of the Lower Saxony Ministry of Science 
and Culture through the ``Nieder\-s\"achsi\-sches Vorab'' grant program 
(grant ZN 2764/ ZN 2896).
The simulations were performed at the HERO cluster of the University
of Oldenburg funded by the DFG (INST 184/108-1 FUGG) and the 
ministry of Science and Culture (MWK) of the Lower Saxony State.

\bibliographystyle{epj}
\bibliography{alex_refs}

\end{document}